\documentclass{aa}  

\usepackage{natbib,twoopt}
\usepackage{multirow}
\usepackage{amsmath}
\usepackage{amssymb} 
\usepackage[T1]{fontenc}
\usepackage{blindtext}
\usepackage{hyperref}
\hypersetup{
    colorlinks=true,
    linkcolor=red,
    filecolor=blue,
    citecolor=blue,
    urlcolor=blue,
    pdftitle={Overleaf Example},
    pdfpagemode=FullScreen,
}
\usepackage{graphicx}
\usepackage{gensymb}
\usepackage{subcaption}
\usepackage{pythonhighlight}
\usepackage{listings}
\usepackage{xcolor}
\usepackage{multirow}
\usepackage{siunitx}
\usepackage{makecell}
\usepackage{orcidlink}
\usepackage{float}

\usepackage{amsmath} 

\usepackage[varg]{txfonts}

\usepackage{graphicx}	
\usepackage{amsmath}	

\definecolor{codegreen}{rgb}{0,0.6,0}
\definecolor{codegray}{rgb}{0.5,0.5,0.5}
\definecolor{codepurple}{rgb}{0.58,0,0.82}
\definecolor{backcolour}{rgb}{0.95,0.95,0.92}

\lstdefinestyle{mystyle}{
  backgroundcolor=\color{backcolour}, commentstyle=\color{codegreen},
  keywordstyle=\color{magenta},
  stringstyle=\color{codepurple},
  basicstyle=\ttfamily\footnotesize,
  breakatwhitespace=false,         
  breaklines=true,                 
  captionpos=b,                    
  keepspaces=true,                                               
  showspaces=false,                
  showstringspaces=false,
  showtabs=false,                  
  tabsize=2
}

\begin{document}

\title{Revisiting the distance to 4U 1543-475 through X-ray dust-scattering rings}
\author{
B. Vaia\orcidlink{0000-0003-0852-0257}\inst{1}
\and
G. Mastroserio\orcidlink{0000-0003-4216-7936}\inst{2}
\and
T. Darie\inst{3}
\and
A. Tiengo\orcidlink{0000-0002-6038-1090}\inst{1,2}
}

\authorrunning{B. Vaia et al.}

\institute{
INAF -- Istituto di Astrofisica Spaziale e Fisica Cosmica di Milano,
via A. Corti 12, 20133 Milano, Italy
\and
Scuola Universitaria Superiore IUSS Pavia, Piazza della Vittoria 15, 27100 Pavia, Italy
\and
Alexandru Ioan Cuza University of Ia\c{s}i, Ia\c{s}i, 700506, Romania
}

\abstract{
We derived a geometrical distance 
of the black-hole X-ray binary 4U~1543$-$475 based on X-ray dust-scattering rings associated with its 1992 and 2002 outbursts. We analyzed archival \textit{ROSAT} and \textit{XMM-Newton} observations and identified an X-ray dust-scattering ring in each observation, produced by two different dust layers along the line of sight. By combining the angular radii of the rings with the independently determined distances of the scattering dust layers and the epochs of the corresponding outbursts, we derive a best-constrained distance to 4U~1543$-$475 of $4.6\pm0.2$~kpc. This value is substantially smaller than the distance of $7.5$~kpc commonly adopted in the literature and has important implications for the inferred physical properties of the system. In particular, the revised distance favors the presence of a low-mass extended-horizontal-branch donor star over a main-sequence star, calling for a reassessment of the dynamical parameters of the binary and, consequently, of the black-hole mass. The smaller distance also affects the inferred properties of the relativistic jet and the estimate of black-hole spin derived from the spectral continuum-fitting.
}

\keywords{}

\maketitle
\nolinenumbers
\section{Introduction}\label{sec1}
4U~1543$-$475 is a Galactic black hole X-ray binary first detected during its 1971 outburst by \textit{Uhuru} \citep{Matilsky1972}. Since its discovery, the source has undergone several recurrent X-ray outbursts, observed in 1983, 1992, 2002, and 2021, in addition to the discovery outburst in 1971. The parameters most commonly adopted in the literature are $i_{\rm orb}=20.7^{\circ}\pm1.5^{\circ}$, derived from the modeling of the ellipsoidal modulation of the companion-star optical light curves \citep{Orosz1998,Orosz2003}. The companion, IL~Lup, is commonly classified as an A2V main sequence star \citep{Orosz1998} with a mass $M_{2} = 2.45\pm 0.15\, M_{\odot}$ \citep{Orosz1998}. A black-hole mass of $M_{\rm BH}=9.4\pm1.0\,M_{\odot}$ is widely adopted in the literature \citep{Ritter2003}.
The distance to 4U~1543$-$475, however, remains comparatively uncertain. A value of $7.5\pm0.5$~kpc, derived from optical observations by combining the spectral classification and photometric properties of the companion star with estimates of the interstellar extinction and the contribution of the accretion disk \citep{Jonker2004}, has been widely adopted in the literature. More recently, estimates incorporating the \textit{Gaia} DR3 parallax together with a Galactic spatial prior for black hole low-mass X-ray binaries favour a smaller distance of $5.0^{+2.0}_{-1.2}$~kpc \citep{Zhang2026}.

An independent way of measuring distances to Galactic X-ray sources is provided by X-ray scattering from interstellar dust. Dust scattering of X-rays emitted by bright point sources is a well-established and extensively investigated phenomenon. At soft X-ray energies, a fraction of the photons emitted by the source is scattered through small angles by dust grains located along the line of sight. Because the scattered photons travel a longer path than the photons reaching the observer directly, they are detected with a time delay.
For persistent sources, the photons scattered by dust distributed along the line of sight produce a diffuse X-ray halo surrounding the source \citep{Premitt,Smith2008}. By contrast, when a sufficiently short X-ray flare illuminates relatively thin dust layers, the delayed scattered emission can appear as one or more discrete rings around the source \citep{Vaughan2004,Tiengo2010,Heinz2016,Pintore2017}. As time passes after the illuminating flare, photons scattered at progressively larger angles reach the observer, producing the characteristic apparent expansion of the rings.
This geometrical dependence makes dust-scattering rings particularly powerful probes of both the distribution of interstellar material and the distances of transient X-ray sources. If the source distance is known or the source is extragalactic, the angular expansion of the rings can be used to determine the distance of the scattering dust layer (e.g. \citealt{Tiengo2023,Vaia2025,Vaia2026b}). Conversely, if the distance of the dust layer can be independently constrained, for example through three-dimensional extinction maps or measurements of interstellar gas, the same geometry can be inverted to determine the distance to the X-ray source. A particularly relevant example for this last case is provided by \citet{Lamer2021}, who used \textit{eROSITA} observations of an expanding dust-scattering ring around the black-hole low-mass X-ray binary MAXI~J1348$-$630 to derive a geometrical distance to the system.

In this work, we present a geometrical distance determination to 4U~1543$-$475 based on the detection of X-ray dust-scattering rings in archival \textit{XMM-Newton} and \textit{ROSAT} observations of its 2002 and 1992 outbursts, respectively.  The paper is organized as follows. Section~\ref{sec2} describes the X-ray observations and data reduction. Section~\ref{sec3} presents the radial 
surface-brightness profiles, the scattering geometry, and the resulting distance measurements. Section~\ref{sec4} discusses the implications of our results, and Section~\ref{sec5} summarizes our conclusions.

\section{Observations and Data reduction}\label{sec2}

\subsection{XMM-Newton}\label{sec2.2}

4U~1543$-$475 was observed with \textit{XMM-Newton} on 2002 August 18, from 12:30 to 20:42 UT (obsID: 0155762201), for a total exposure of approximately 29.5 ks. During the observation, all three \textit{XMM-Newton}/EPIC detectors were active \citep{MOScamera,PNcamera}. EPIC-pn was operated in Small Window mode, while MOS1 and MOS2 were configured in Timing Uncompressed and Full Frame modes, respectively. A Thin1 optical blocking filter was used for each detector. 
For our analysis, we considered the data from the EPIC MOS2 camera, which was the only one covering the full field of view. 
The event file was processed with the \textit{XMM-Newton} Science Analysis 
Software (SAS; version 20.0.0; \citealt{2004ASAS}), using the most recent 
calibration files available. Standard screening criteria were applied to 
select valid MOS events\footnote{\texttt{\#XMMEA\_EM\&\&(PATTERN<=12)}}. 
Point sources detected with the SAS task \texttt{emldetect} were removed 
from the analysis by excluding circular regions with a radius of 
$25^{\prime\prime}$ around their positions.

\begin{figure}
    \centering
    \includegraphics[width=0.8\columnwidth]{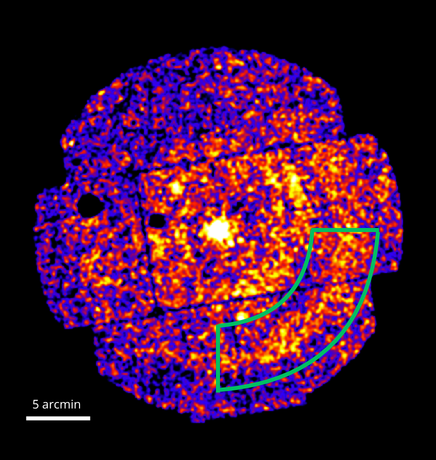}
    \caption{\textit{XMM-Newton}/EPIC-MOS2 image of 4U~1543$-$475 in the 0.5--2 keV energy band. Point sources in the field have been excluded. The green annular sector indicates the region where the fit shown in Fig.~\ref{rp_xmm} was performed.}
    \label{halo_fig}
\end{figure}

For the analysis, we selected the $0.5$--$2$ keV energy band, which 
maximizes the signal-to-noise ratio of the dust-scattering ring. 
The background contribution was instead estimated using the high-energy 
$5$--$12$ keV band, as in \citet{Vaia2025,Vaia2026b}. Furthermore, based 
on a visual inspection of the azimuthal distribution of the emission, 
we restricted the analysis to the $270^{\circ}$--$360^{\circ}$ sector, 
where the dust-scattering signal is most prominent (see Fig. \ref{halo_fig}), thereby further 
improving the signal-to-noise ratio.

\subsection{ROSAT} \label{sec2.1}

The \textit{ROSAT} satellite observed 4U~1543$-$475 on 1992 August 21 with the Position Sensitive Proportional Counter (PSPC), which operates in the 0.1--2.4 keV energy range. During the observation with a net exposure of 1.4 ks, the source had a mean count rate of about 1000 cts s$^{-1}$. As in the \textit{XMM-Newton} analysis, we restricted the \textit{ROSAT} data to the 0.5--2 keV energy band. An exposure map was created to account for the spatially variable efficiency of the mirrors and PSPC detector.

\section{Data analysis and results}\label{sec3}

\subsection{Radial Profile of the observations}\label{sec3.1}

\begin{figure}
    \centering
    \includegraphics[width=\columnwidth]{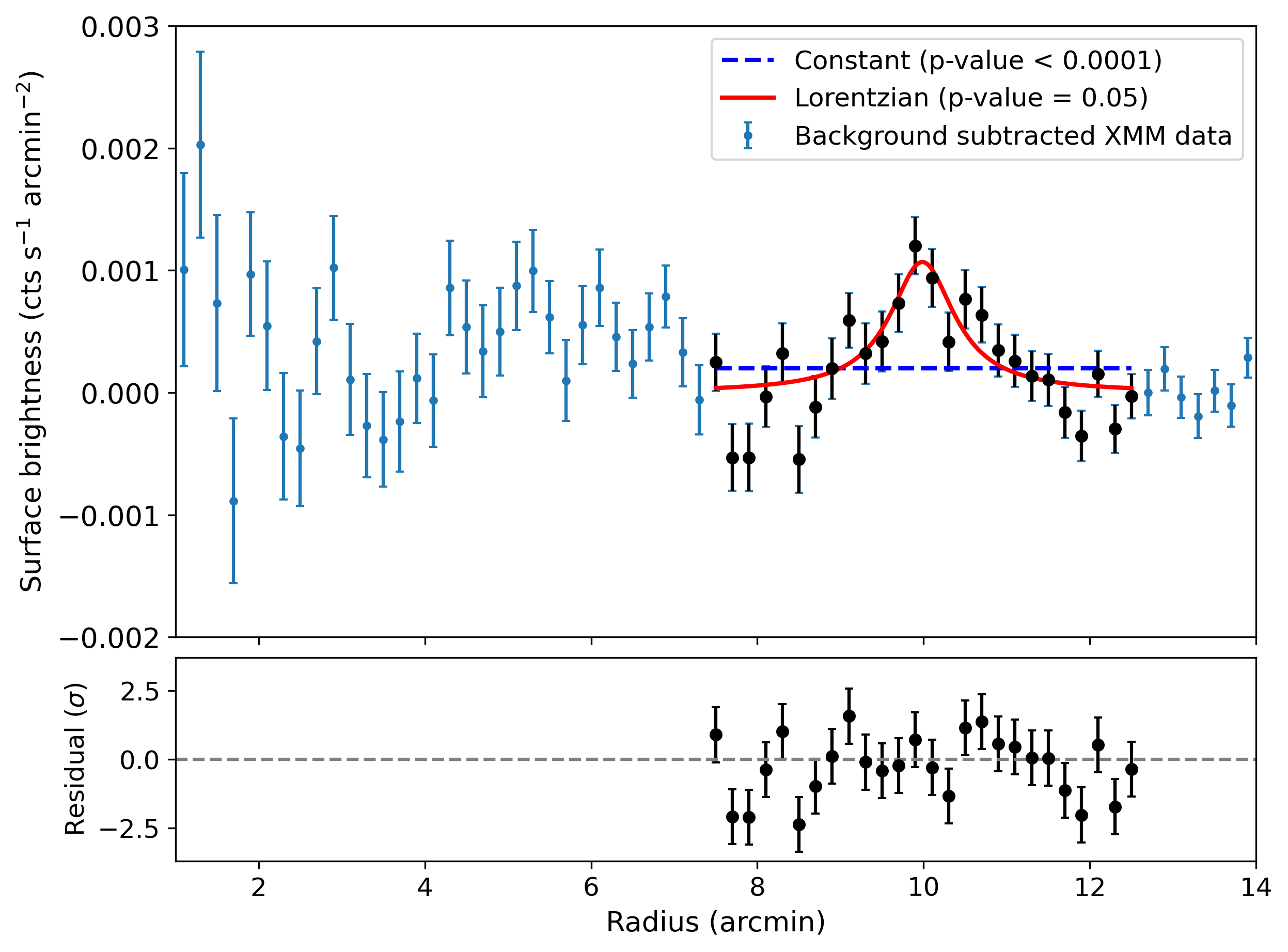}
    \caption{Background subtracted radial surface-brightness profile of the \textit{XMM-Newton} MOS2
    observation of 4U~1543$-$475 in the 0.5--2 keV energy band, extracted 
    within the $270^\circ$--$360^\circ$ sector (blue points).
    The fit was performed over the $\theta=7.5$--$12.5$ arcmin radial range (black points). 
    The blue dashed line represents the best-fitting constant model, 
    corresponding to the null hypothesis of no ring-like excess, while the 
    red solid line shows the best-fitting Lorentzian model. The lower panel shows the residuals in units of the statistical uncertainty.}
    \label{rp_xmm}
\end{figure}

\begin{figure}
    \centering
    \includegraphics[width=\columnwidth]{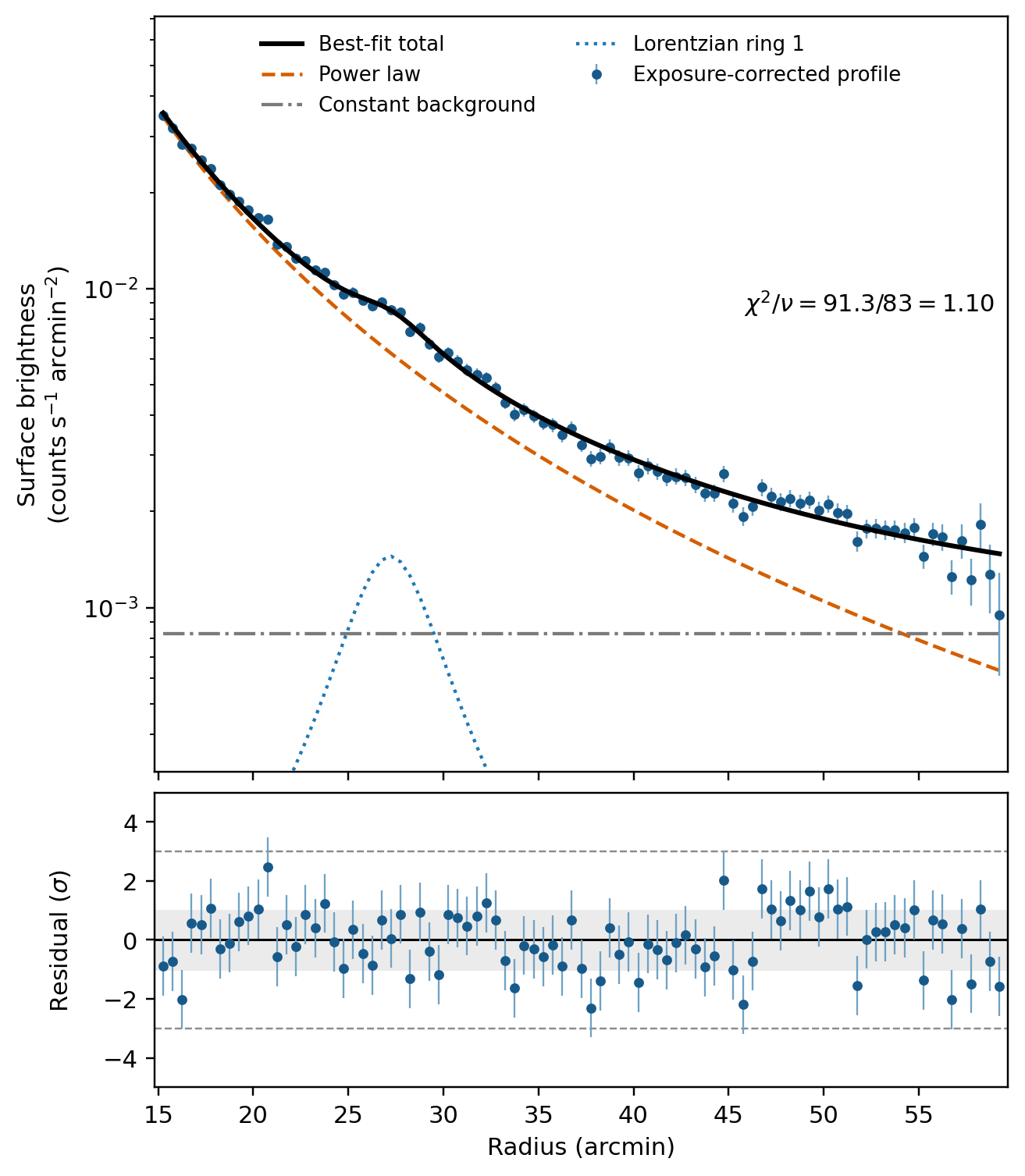}
    \caption{
    Radial surface-brightness profile of the ROSAT observation of 4U~1543$-$475.
    The blue points show the exposure-corrected radial profile, while the solid black line represents the best-fitting model, consisting of a power-law component (orange dashed line), a constant background (gray dash-dotted line), and a Lorentzian component describing the dust-scattering ring (blue dotted line).
    The best fit gives $\chi^2/\nu = 91.3/83 = 1.10$.
    The lower panel shows the residuals in units of the statistical uncertainty, with the dashed horizontal lines indicating $\pm3\sigma$.
    }
    \label{rp_rosat}
\end{figure}

To characterize the diffuse X-ray emission surrounding 4U~1543$-$475 in the two observations described in Sect.~\ref{sec2}, we extracted radial surface-brightness profiles centered on the source position from the \textit{XMM-Newton} (Fig.~\ref{rp_xmm}) and \textit{ROSAT} (Fig. \ref{rp_rosat}) observations.

For the \textit{XMM-Newton} observation, we constructed the radial surface-brightness profile within the selected $270^\circ$--$360^\circ$ sector (measured counterclockwise from West, see Fig. \ref{halo_fig}). A radial profile was also
derived from the background data, as described in Sect.~\ref{sec2.2}. Prior to subtraction, the background
profile was rescaled to match the observed profile within a radial interval free from any detectable contribution from the dust-scattering rings. The resulting normalized background profile was then subtracted from the observed one to obtain the background-subtracted radial profile used in the subsequent analysis.
The resulting profile reveals a well-defined ring-like excess at approximately $10$ arcmin, together with an
additional, broader excess at $\sim5$ arcmin. Although the latter feature may also be associated with
dust-scattered emission, its broad radial extent prevents an accurate determination of its characteristic
angular radius and, consequently, does not provide a sufficiently robust constraint for the geometrical
analysis.
We therefore restricted the fit to the radial interval $\theta=[7.5,12.5]$ arcmin, encompassing the narrower
and better-defined ring-like feature. Within this interval, we first fitted the profile with a constant model, representing the null hypothesis of no ring-like excess, and subsequently with a Lorentzian model to
characterize the ring feature \citep{Tienghetti2006}.
The Lorentzian model provides a significantly better description of the data than the constant model, improving the fit from $\chi^2/\nu=91/25$ to $\chi^2/\nu=35/23$ (p-value = 0.05\footnote{We assess the goodness of fit by calculating the p-value from the $\chi^2$ distribution, assuming as the null hypothesis that the model provides an adequate description of the data. We adopt a significance threshold of 0.01, rejecting models with p-values below this limit.}). The best-fitting Lorentzian centroid yields a ring angular radius of $\theta=9.98\pm0.07$ arcmin.

For the \textit{ROSAT} observation, we fitted the exposure-corrected radial surface-brightness profile over the radial range $\theta=[15,60]$ arcmin. We first considered a baseline model consisting of a constant, to model a uniform background, plus a power-law component, accounting for the tail of the point-spread function (PSF) and the dust halo of the central point source. We then added a Lorentzian component to describe the excess associated with a possible dust-scattering ring.
The inclusion of the Lorentzian component substantially improves the fit with respect to the baseline model,
reducing $\chi^2/\nu$ from $160/86$ to $91/83$ (p-value = 0.25). The best-fitting Lorentzian centroid yields a ring angular radius of $\theta=27.2\pm0.3$ arcmin. 

In both observations, the centroid of the Lorentzian component provides the angular radius of the scattering ring, whereas its width reflects the combined effects of the instrumental PSF, the
finite duration of the source outburst, and the physical thickness of the scattering dust layer.
These measurements of the ring angular radii provide the basis for the geometrical analysis presented in
the following sections.

\subsection{The scattering geometry}\label{sec3.2}

The angular radius of an expanding ring, $\theta$, is described by:
\begin{equation}
\label{eq_theta1}
\theta = \sqrt{\frac{2c \Delta t}{D_d} \left(1 - \frac{D_d}{D_s}\right)},
\end{equation}
where $c$ is the speed of light, $\Delta t$ represents the time delay between the direct and scattered X-ray photons, $D_s$ is the distance to the X-ray source, and $D_d$ is the distance to the intervening dust cloud.
Therefore, if the location of the scattering dust along the line of sight is known, dust-scattering X-ray observations can be used to directly constrain the distance to the source \citep{Miralda1999}. 
Since the angular radii of the rings, $\theta$, have been measured from the radial profiles in Sect.~\ref{sec3.1}, the remaining quantities required to constrain $D_s$ are the appropriate time delay, $\Delta t$, and the distance to the scattering dust layer, $D_d$. These quantities are determined in the following two sections from the soft X-ray light curve of the outburst and from the distribution of interstellar dust along the line of sight toward 4U~1543$-$475, respectively.

\subsection{Light curve of the flare X-ray emission}\label{sec3.3}
The emission responsible for producing the dust-scattering rings is the soft X-ray emission released during the source outbursts. Indeed, the dust-scattering cross section strongly increases toward lower photon energies, making the soft X-ray emission particularly relevant for the observed scattered signal. In our case, the rings detected in the \textit{ROSAT} and \textit{XMM-Newton} observations are associated with two distinct outbursts of 4U 1543$-$475, which occurred in 1992 and 2002, respectively. Therefore, to characterize the scattering geometry, we first need to identify, for each outburst, the epoch of the soft X-ray emission responsible for producing the corresponding rings.

\subsubsection{The 2002 outburst of 4U~1543$-$475} 
\begin{figure}
\centering
\includegraphics[width=\linewidth]{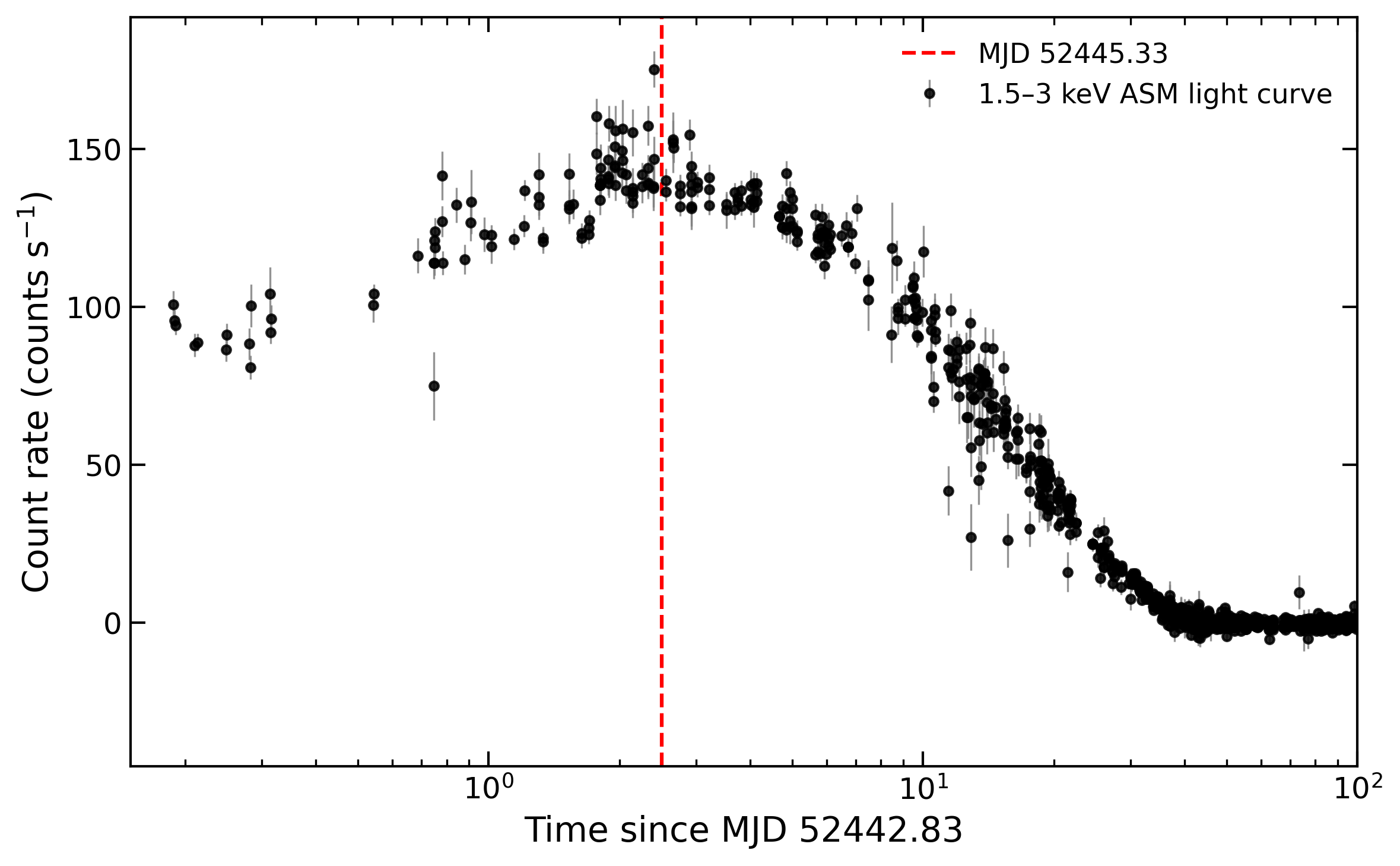}
\caption{RXTE/ASM light curve of the 2002 outburst of 4U~1543$-$475 in the 1.5--3~keV energy band. Time is expressed in days since MJD~52442.83 and shown on a logarithmic scale to better highlight the early phases of the outburst. The red dashed vertical line marks MJD~52445.33 and indicates the maximum of the light curve.}
\label{lc2002} 
\end{figure}

The 2002 outburst is particularly well suited for this analysis, as its entire temporal evolution was monitored by the All-Sky Monitor (ASM, \citealt{Levine1996}) on board the \textit{RossiXTE} satellite. In particular, its 1.5--3 keV light curve closely matches the energy range where the dust-scattering signal was observed by \textit{XMM-Newton}. From the ASM light curve shown in Fig. \ref{lc2002} we derive MJD~52445.33 as the time of the outburst peak and adopt this epoch as the reference time of the photons producing the ring. Since the illuminating emission is not instantaneous, we conservatively adopt an uncertainty of \(\pm3\) days on this reference epoch, approximately accounting for the temporal extent of the bright soft X-ray emission around the maximum.
We note that during the 2002 outburst, 4U~1543$-$475 was initially in the hard state during the rapid rise, before undergoing a fast hard-to-soft state transition close to the peak of the X-ray light curve \citep{Park2004}.

\subsubsection{The 1992 outburst of 4U~1543$-$475}

The situation is less straightforward for the 1992 outburst. Unlike the 2002 event, the 1992 outburst was covered only in the hard X-ray band by the Burst and Transient Source Experiment (BATSE, \citealt{Fishman1985}) on board the \textit{Compton Gamma Ray Observatory} (\textit{CGRO}). Therefore, no continuous soft X-ray light curve is available to directly trace the emission most relevant for dust scattering. Nevertheless, the BATSE observations provide the only available constraint on the timing of the main emission episode, with the peak reported at MJD $\sim$48732 (1992 April 20; \citealt{Harmon1992}).
During all the other outbursts, 4U~1543-475 displayed a typical soft X-ray transient behaviour, which is characterised by very short hard-state raise and hard-to-soft transition, to show then a more typical behavior during the disk dominated soft state \citep{Tetarenko2016}.
Assuming that the 1992 outburst underwent a spectral evolution similar to that observed in the other outbursts, we adopt the epoch of the BATSE peak as a proxy for the time of maximum soft X-ray emission responsible for the dust-scattering feature detected by \textit{ROSAT}. Since the soft X-ray evolution of the 1992 outburst was not directly monitored, the illuminating epoch is considerably less well constrained than for the 2002 outburst. We therefore conservatively adopt an uncertainty of \(\pm10\) days on the reference epoch and propagate it into the distance determination. We stress, however, that this uncertainty is not directly constrained by soft X-ray observations and it may underestimate the actual uncertainty if the spectral evolution of the 1992 outburst differed significantly from that observed in the rest of the source outbursts. In particular, we cannot exclude the possibility that the observed ring was produced by a later
episode of soft X-ray emission, possibly associated with the secondary hard X-ray maximum reported by \citet{darkweb}, approximately 60--110 days after the initial outburst. The implications of this alternative scenario are discussed in Sect.~\ref{ROS_d}.

\subsection{Dust along the line of sight of 4U1543-475}\label{sec3.4}

\begin{figure}
\centering
\includegraphics[width=\linewidth]{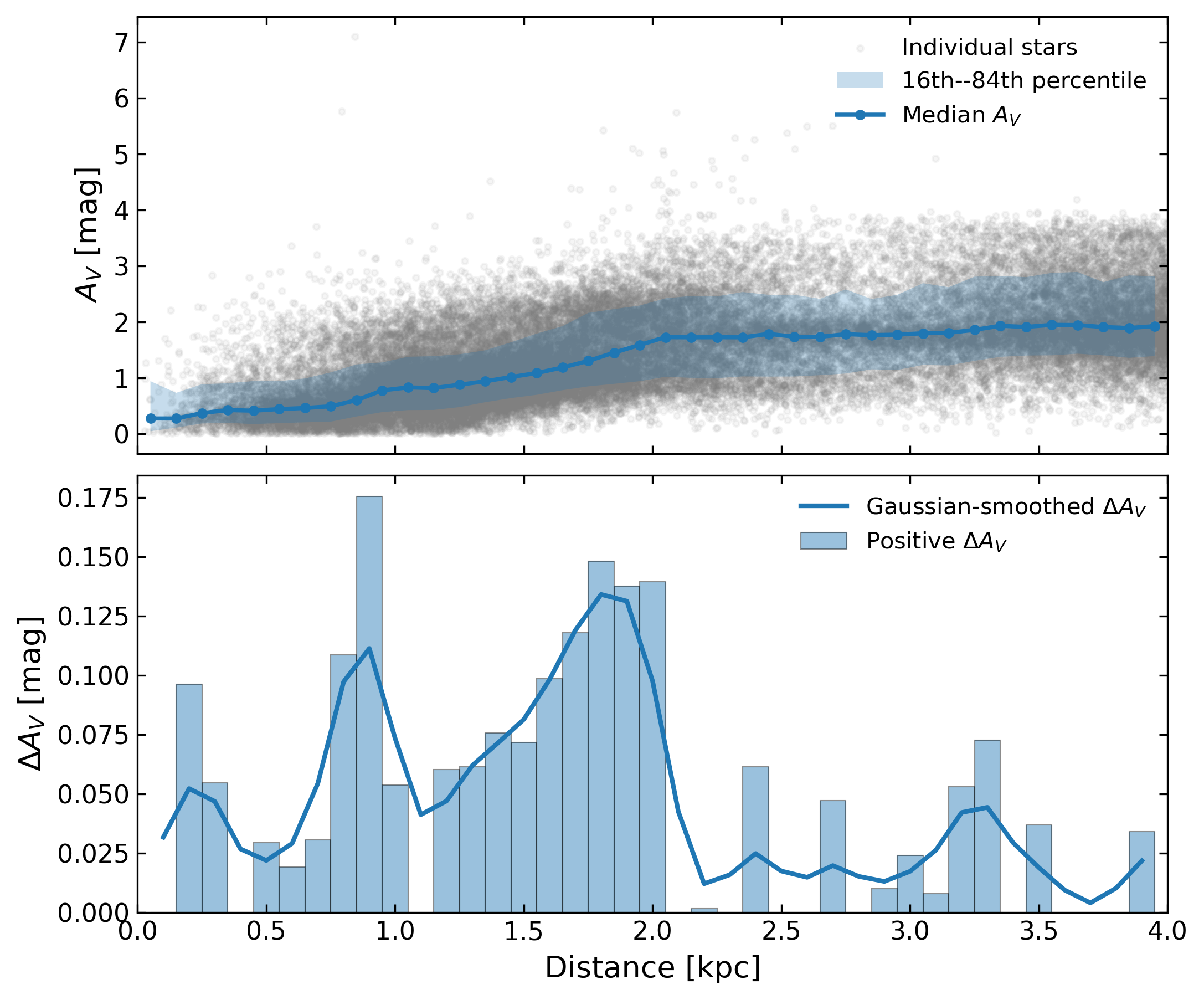}
\caption{Extinction distribution along the line of sight toward 4U 1543-475 derived from the StarHorse catalogue. Top panel: Individual stellar extinction measurements ($A_V$) as a function of distance. The solid line shows the median $A_V$ computed in 0.1 kpc distance bins, while the shaded region represents the 16th--84th percentile range. Bottom panel: Differential extinction ($\Delta A_V$) obtained from the variation of the median cumulative extinction between consecutive distance bins. The solid curve represents the Gaussian-smoothed profile, adopting a Gaussian kernel.}
\label{starhorse_los} 
\end{figure}

To identify the distance of the dust clouds potentially responsible for the observed X-ray scattering, we investigated the distribution of interstellar dust along the line of sight toward 4U~1543$-$475. For this purpose, we used the Gaia DR2 StarHorse catalogue \citep{Anders2019}, which provides distances, extinctions, and other stellar parameters for approximately 265 million stars. These parameters were inferred with the StarHorse Bayesian code \citep{Queiroz2018} by combining Gaia DR2 data with Pan-STARRS1, 2MASS, and AllWISE photometry.
We selected all stars within a $1^{\circ}$ radius centered on the position of 4U~1543$-$475. The resulting extinction distribution is shown in Fig.~\ref{starhorse_los}. Three increases in extinction are visible at distances of approximately $0.9$, $1.8$, and $3.3$~kpc, indicating the presence of distinct dust structures
along the line of sight.
We focused our analysis on the dust structures at $1.8$ and $3.3$~kpc, which are subsequently associated
with the observed X-ray scattering rings (see Section \ref{sec3.5}).
To determine the distances of these two dust clouds and their associated uncertainties, we applied a bootstrap procedure to the StarHorse data, accounting for both statistical fluctuations in the stellar sample and the dependence of the extinction profile on the adopted distance binning 
(see Appendix~\ref{app:bootstrap} for details).

The resulting bootstrap distributions identify two distinct dust clouds along the line of sight, located at $D_{\rm d,1}=1.84\pm0.08$~kpc and $D_{\rm d,2}=3.26\pm0.08$~kpc.
The positions of both extinction enhancements remain stable against variations in the adopted bin width and bin offset.

As an independent check, we also investigated the distribution of molecular gas traced by Galactic CO emission and the three-dimensional dust distribution derived from stellar extinction maps \citep{Lallement2022}. Both provide a qualitatively consistent picture of the main structures identified with StarHorse. These complementary analyses are presented in Appendix~\ref{app:dust_los}.

\subsection{Distance determination}\label{sec3.5}

Having independently constrained the angular radius of the scattering rings, the epoch of the illuminating X-ray emission, and the distances of the dust clouds along the line of sight, we can now use the scattering geometry to determine the distance to 4U~1543$-$475. Rearranging Eq.~\ref{eq_theta1}, the source distance can be expressed as \begin{equation} \label{eq_distance}
D_s = \frac{D_d} {1-\dfrac{\theta^2 D_d}{2c\Delta t}}, 
\end{equation} 
where $\Delta t$ is calculated as the time interval between the adopted epoch of the illuminating flare and the corresponding X-ray observation.

\subsubsection{\textit{XMM-Newton} distance constraint}

Following Eq. \ref{eq_distance}, if we associate the ring detected at $\theta=9.98\pm0.07$ arcmin with the dust cloud at $D_{\rm d,2}=3.26\pm0.08$ kpc, together with the epoch of the 2002 soft X-ray peak derived in Sect.~\ref{sec3.3}, we obtain a distance to 4U~1543$-$475 of:
\begin{equation}
D_s = 4.6\pm0.2~{\rm kpc}.
\end{equation}
Associating the observed ring with the closer dust layer at $\sim1.8$ kpc would instead imply a substantially smaller source distance of 2.1 kpc, which would be even
smaller if the ring were associated with the dust structure at $\sim0.9$ kpc. Both scenarios are
incompatible with the independent distance constraint provided by the Gaia parallax
($5.0_{-1.2}^{+2.0}$ kpc, \citealt{Zhang2026}).
If the X-ray ring detected at $\sim$10 arcmin is due to dust at $\sim$3.3 kpc,
the scattering ring produced by the closer dust cloud at $\sim1.8$ kpc is expected at a substantially larger angular radius. Such a ring would fall outside the \textit{XMM-Newton} field of view, naturally explaining why only the ring associated with the more distant dust layer is detected in this observation.
\subsubsection{\textit{ROSAT} distance constraint}\label{ROS_d}

The larger field of view of \textit{ROSAT} allows us to probe scattering structures at larger angular radii. We therefore associate the ring detected at $\sim27$ arcmin with the closer dust layer at $D_{\rm d,1}=1.84\pm0.08$ kpc. This association is also supported by the StarHorse extinction profile, which shows a stronger increase in $A_V$ at this distance, indicating a larger amount of scattering material along the line of sight. Adopting the epoch of the 1992 outburst discussed in Sect.~\ref{sec3.3} and applying Eq.~\ref{eq_distance}, we obtain an independent estimate of the source distance of:
\begin{equation}
D_s = 4\pm1~{\rm kpc}.
\end{equation}
In this case, a ring produced by the more distant dust layer at $\sim3.3$ kpc would appear at a smaller angular radius, making its detection challenging due to contamination from the bright point source. Conversely, a ring associated with the dust structure at $\sim0.9$ kpc would appear at larger angular radii, where dust scattering becomes less efficient.

We also explored the possibility that the observed ring was produced by the second peak detected by BATSE, considering the different dust layers along the line of sight. The only combination yielding a physically plausible source distance is the one described above. An overview of the different combinations considered and the corresponding source distances is provided in Table~\ref{tab}.

\begin{table}
\centering
\caption{Source distances inferred from Eq. \ref{eq_distance} for $\theta=27'$ (corresponding to the ROSAT ring) and for different dust-layer distances and time delays.}
\label{tab}
\begin{tabular}{cccc}
\hline
$d_{\rm dust}$ & $\Delta t=15$ d & $\Delta t=65$ d & $\Delta t=123$ d \\
\hline
0.9 kpc & -- & 1.83 kpc & 1.23 kpc \\
1.8 kpc & -- & -- & 4.00 kpc \\
3.3 kpc & -- & -- & 230 kpc \\
\hline
\end{tabular}
\tablefoot{The tabulated values are the inferred source distances in kpc. Dashes indicate combinations for which no physical solution with $d_{\rm source}>d_{\rm dust}$ is obtained.}
\end{table}

\section{Discussion}\label{sec4}

The analysis presented in this work provides a new constraint on the distance to 4U~1543$-$475. We obtained a best distance of $D_s=4.6\pm0.2$~kpc studying the \textit{XMM-Newton} dust scattering ring around the 2002 outburst. Remarkably, the independent analysis of the dust-scattering feature detected with \textit{ROSAT} following the 1992 outburst yields $D_s=4\pm1$~kpc. Although the latter estimate is affected by a larger uncertainty because of the lack of direct soft X-ray monitoring of the 1992 outburst, which prevents us to precisely determine the emission peak, the agreement between the two measurements provides important support for the proposed distance estimate.

\subsection{Comparison with previous distance estimates}

The first detailed distance estimate for 4U~1543$-$475 was provided by \citet{Orosz1998}, who derived a value of approximately $9.1$~kpc. A distance of $7.5\pm0.5$~kpc was later quoted by \citet{Jonker2004} and has subsequently become the value most commonly adopted in the literature. Interestingly, \citet{Jonker2004} attributed this estimate to \citet{Orosz2002}; however, the latter is a conference abstract and does not explicitly report a distance of $7.5$~kpc. The origin of the widely adopted $7.5$~kpc value is therefore not clearly documented in the cited literature. 
Our result instead places 4U~1543$-$475 substantially closer. A smaller distance has also been independently suggested by recent estimates based on \textit{Gaia} astrometry. In particular, \citet{Zhang2026} combined the \textit{Gaia} DR3 parallax ($0.191\pm0.055$ mas) with a prior based on the Galactic distribution of black-hole low-mass X-ray binaries \citep{Atri2019}, obtaining a distance of $5.0^{+2.0}_{-1.2}$ kpc. Although considerably less precise than our geometrical determination, this estimate is consistent with our values of $4.6\pm0.2$~kpc (\textit{XMM-Newton)} and $4\pm1$~kpc (\textit{ROSAT}), and independently favors a distance smaller than the most commonly adopted value of 7.5~kpc.
For the rest of the discussion we are going to assume the most constraining measurement of $4.6\pm0.2$~kpc as the result of our analysis. 

\subsection{Revisiting the binary system parameters}

The distance and the dynamical parameters of the binary system are closely connected through the assumed properties of the donor star. In their optical study, \citet{Orosz1998} classified the companion as an early A-type star with $T_{\rm eff}\simeq9000$~K. However, they explicitly considered two
possible evolutionary interpretations for the donor. If the star is on the main sequence, its expected mass is $\sim2.2\,M_\odot$, whereas an
alternative solution is a highly evolved star on the extended horizontal branch (EHB), with a helium-burning core, a thin hydrogen
envelope, and a mass of $\sim0.5\,M_\odot$. The available spectroscopy could not distinguish between these possibilities through the surface
gravity. \citet{Orosz1998} favored the main-sequence interpretation on evolutionary grounds. They therefore adopted $2.3\leq M_2/M_\odot\leq2.6$, where $M_2$ denotes the mass of the donor star.
Importantly, the two evolutionary interpretations lead to substantially different distance estimates. For the main-sequence
solution, with $R_2=2.84\pm0.11\,R_\odot$ ($R_2$ denotes the donor-star radius), \citet{Orosz1998} obtained $D_s=9.1\pm1.1$ kpc for $E(B-V)=0.50\pm0.05$, or $7.2\pm1.0$ kpc when
adopting $E(B-V)=0.66\pm0.10$\footnote{$E(B-V)$ is the color excess, defined as the difference between the observed and intrinsic $B-V$ color, where $B$ and $V$ refer to the blue and visual bands of the Johnson photometric system.}. Conversely, for the EHB solution with $M_2=0.5\,M_\odot$ and $R_2=1.65\pm0.07\,R_\odot$, they obtained $D_s=5.3\pm0.6$ and $4.2\pm0.6$ kpc for the same two reddening assumptions, respectively.

Our geometrical distance of $4.6\pm0.2$ kpc is therefore inconsistent with the main-sequence solution favored by \citet{Orosz1998}, while it
falls within the range predicted by their alternative low-mass EHB scenario and this have important implication on the compact-object mass ($M_{BH}$).  
Indeed, the compact-object mass is related to the mass function,
inclination, and donor mass through the equation:

\begin{equation}
f(M)=\frac{M_{\rm BH}^{3}\sin^{3}i}
{(M_{\rm BH}+M_2)^2}.
\end{equation}

\citet{Orosz1998} measured $f(M)=0.22\pm0.02\,M_\odot$ and obtained conservative inclination limits of $20^\circ\leq i\leq40^\circ$.
Using the full $3\sigma$ mass-function interval adopted by \citet{Orosz1998}, $0.16\leq f(M)/M_\odot\leq0.28$, together with their conservative inclination limits, and dopting the $M_2\simeq0.5\,M_\odot$ EHB solution we obtain 

\begin{equation}\label{BH_mass0}
1.2\lesssim M_{\rm BH}/M_\odot\lesssim7.9.
\end{equation}

Using, indeed, the values commonly adopted in the literature, i.e. $i=20.7^{\degree}\pm1.5^{\degree}$ and  $f(M)= 0.25\pm0.01M_\odot$ yields a nominal compact-object mass of  $M_{\rm BH}\simeq6.6 M_{\odot}$. 
Accounting for the uncertainties on both parameters, the corresponding 1 $\sigma$ mass range is 

\begin{equation} \label{BH_mass}
5.3\lesssim M_{\rm BH}/M_\odot\lesssim8.2.
\end{equation}



However, the original derivation of these updated values for 
$i$ and $f(M)$ is not readily traceable in the published literature. These updated binary parameters were reported in \citet{Orosz2003} referring to a subsequent dynamical study by Orosz et al., which was cited as being in preparation but, to our knowledge, was never published. Therefore, although these values have subsequently been widely adopted, the analysis from which they were derived is not available in a refereed publication.

\subsection{Implications for the super-Eddington interpretation of the 2021 outburst}

The exceptionally bright 2021 outburst of 4U~1543$-$475 has been interpreted as an episode of super-Eddington accretion. \citet{Jin2024},
assuming a black-hole mass of $9.4\,M_{\odot}$ and a distance of $7.5$~kpc, inferred a peak luminosity of approximately $2.3\,L_{\rm Edd}$ from the spectral evolution observed with \textit{Insight-HXMT} \citep{Zhang2020}. Evidence for super-Eddington accretion was also reported by \citet{Zhao2024} from quasi-simultaneous \textit{Swift}/XRT \citep{Gehrels2004,XRT} and \textit{NuSTAR} \citep{Nustar2013} observations obtained during the subsequent evolution
of the outburst.

The distance derived in this work calls for a reassessment of this interpretation. A smaller distance directly implies a lower intrinsic
luminosity for a given observed flux and would therefore reduce the Eddington ratios inferred in previous studies. Adopting a black-hole mass of 9.4 $M_{\odot}$ and a distance of 4.6 kpc, the Eddington ratio inferred at the peak by \citet{Jin2024} decreases to $\simeq$ 0.87.
However, the effect of
the revised distance cannot be considered independently of the dynamical parameters of the binary. As discussed above, our distance
favors the low-mass EHB interpretation of the donor considered by \citet{Orosz1998}, which would in turn imply a revision of the
black-hole mass. Using the black-hole mass constraints reported in Eq.~\ref{BH_mass0} and the new distance derived in this work, the Eddington ratio of the 2021 outburst can be reassessed. Since $L/L_{\rm Edd} \propto D^2/M_{\rm BH}$, rescaling the values reported in previous studies to our new values yields a peak Eddington ratio of $L/L_{\rm Edd} \simeq 1.0$ -- $6.8$ for the \citet{Jin2024} observation.


\subsection{Implications for the relativistic jet}

The revised distance has also implications for the recently reported relativistic jet ejections from 4U~1543$-$475 \citep{Zhang2026}. By monitoring the 2021 outburst at radio wavelengths with MeerKAT, they inferred a likely Lorentz factor of $\Gamma\sim8$, with a minimum value of $\Gamma>4.6$ at 95\% confidence.
To estimate the Lorentz factor distribution, they put a prior on the distance of $5.0^{+2.0}_{-1.2}$~kpc and run MCMC. 
The more accurate source distance reported here ($4.6\pm0.2$~kpc) allows us to better constrain the Lorentz factor, which, being proportional to the source distance, would be slightly smaller.  
Nevertheless, since the lower limit of the distance derived from our analysis is larger than that assumed by \citet{Zhang2026}, the jet would remain highly relativistic, and, therefore, their main conclusion 
would not be affected.

\subsection{Black--hole spin measurements}

Two main techniques have been used to constrain the spin of 4U~1543$-$475: continuum fitting and
X-ray reflection spectroscopy. While reflection-based measurements do not directly depend on the source distance, the continuum-fitting method requires independent estimates of the black-hole mass, inclination, and distance.
Previous continuum-fitting analyses of 4U~1543$-$475, generally adopting a distance of $\sim7.5$~kpc and a black-hole mass of 9.4 $M_{\odot}$, struggled to reach a consistent spin measurement: $0.43^{+0.22}_{-0.31}$ \citep{Miller2009,Morningstar2014}, $0.65^{+0.14}_{-0.24}$ \citep{Yorgancioglu2023}, and $\sim0.75-0.85$ \citep{Shafee2006}.
A smaller distance implies a smaller inferred inner disk radius and, assuming that the disk extends down to the ISCO, a higher black-hole spin. The substantially smaller distance of $4.6\pm0.2$~kpc derived in this work would thus favor a higher spin than previously inferred from continuum-fitting measurements. However, the revised distance also motivates a reassessment of the black-hole mass, as discussed above. Since a lower black-hole mass would lead to a larger inner disk radius in gravitational units, it would instead favor a lower spin. These competing effects prevent us from drawing a firm conclusion about the impact of the revised distance on the inferred black-hole spin.
A reanalysis of the continuum spectrum accounting for both the revised distance and the associated uncertainties in the black-hole mass is beyond the scope of this work.

\section{Conclusion and future prospectes}\label{sec5}

We presented a new geometrical distance determination for the Galactic black hole X-ray binary 4U~1543$-$475, based on the analysis of X-ray dust-scattering rings detected in archival \textit{XMM-Newton} and \textit{ROSAT} observations of the 2002 and 1992 outbursts, respectively.
We obtained two independent distance estimates: $D_s = 4.6 \pm 0.2$ kpc from the \textit{XMM-Newton} observation, and $D_s = 4 \pm 1$ kpc from the \textit{ROSAT} observation. The agreement between these two independent measurements, obtained from different outbursts, different instruments, and different dust structures along the line of sight, provides strong support for the revised distance.
Our result places 4U~1543$-$475 substantially closer than the value of $7.5 \pm 0.5$ kpc commonly adopted in the literature, and is consistent with the independent, though less accurate, \textit{Gaia} DR3-based estimate of $5.0^{+2.0}_{-1.2}$ kpc \citep{Zhang2026}. 
The revised distance has direct consequences for the physical interpretation of the source. In particular, the distance derived here is difficult to
reconcile with the main-sequence donor favored by \citet{Orosz1998}, while it is consistent with the alternative low-mass extended-horizontal-branch (EHB) solution considered in the same work. Since the inferred mass of the compact object depends on the assumed donor mass and orbital inclination, a new dynamical analysis of the system is required to obtain a self-consistent determination of its component masses. Adopting the EHB donor solution and the dynamical parameters reported in \citet{Orosz1998} we obtain a black-hole mass range of $1.2\lesssim M_{\rm BH}/M_\odot\lesssim7.9$. 
The jet Lorentz factor inferred by \citet{Zhang2026} shifts toward the lower end of its allowed range, while remaining highly relativistic. The black-hole spin inferred from continuum-fitting spectroscopy may instead be affected by the competing effects of the revised distance and black-hole mass.

More broadly, this work illustrates how powerful X-ray dust-scattering echoes are as geometrical distance estimators for Galactic X-ray binaries. The present analysis relied entirely on archival \textit{XMM-Newton} and \textit{ROSAT} observations obtained more than two decades ago, neither of which was originally intended to characterize dust scattering. This underscores two broader points: 
\begin{itemize}
    \item The X-ray archives of \textit{XMM-Newton}, \textit{Chandra}, \textit{ROSAT}, and \textit{Swift}/XRT contain decades of pointed observations of Galactic X-ray transients following bright outbursts, many of which have never been systematically searched for dust-scattering rings and can lead to valuable results.
    \item New dedicated or serendipitous observations with present or future imaging x-ray instruments will be a powerful tool to measure the distance of Galactic transients without a well determined \textit{Gaia} parallax.
\end{itemize}

\appendix

\section{The bootstrap algorithm}
\label{app:bootstrap}

\begin{figure*}
    \centering
    \includegraphics[width=0.48\textwidth]{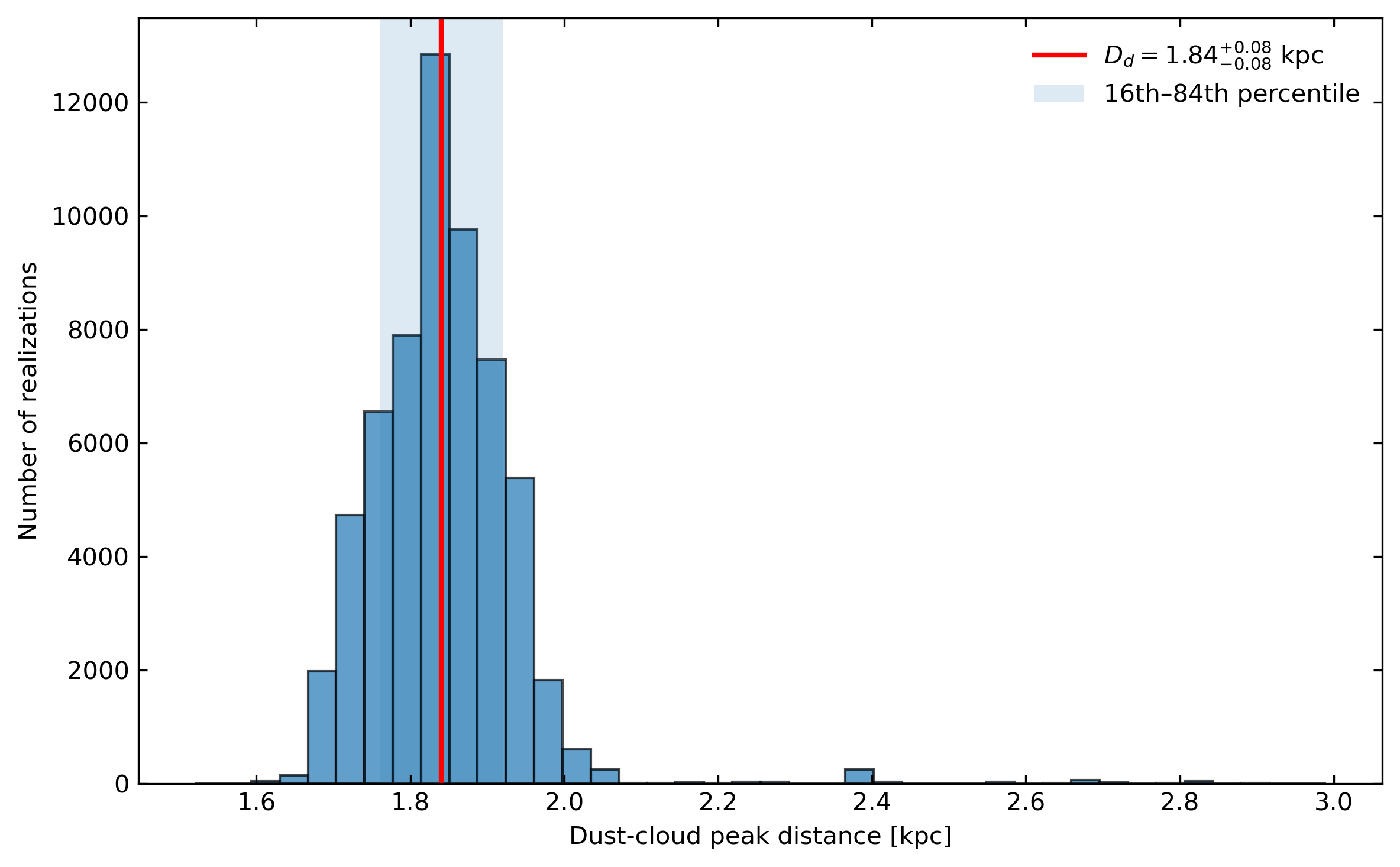}
    \hfill
    \includegraphics[width=0.48\textwidth]{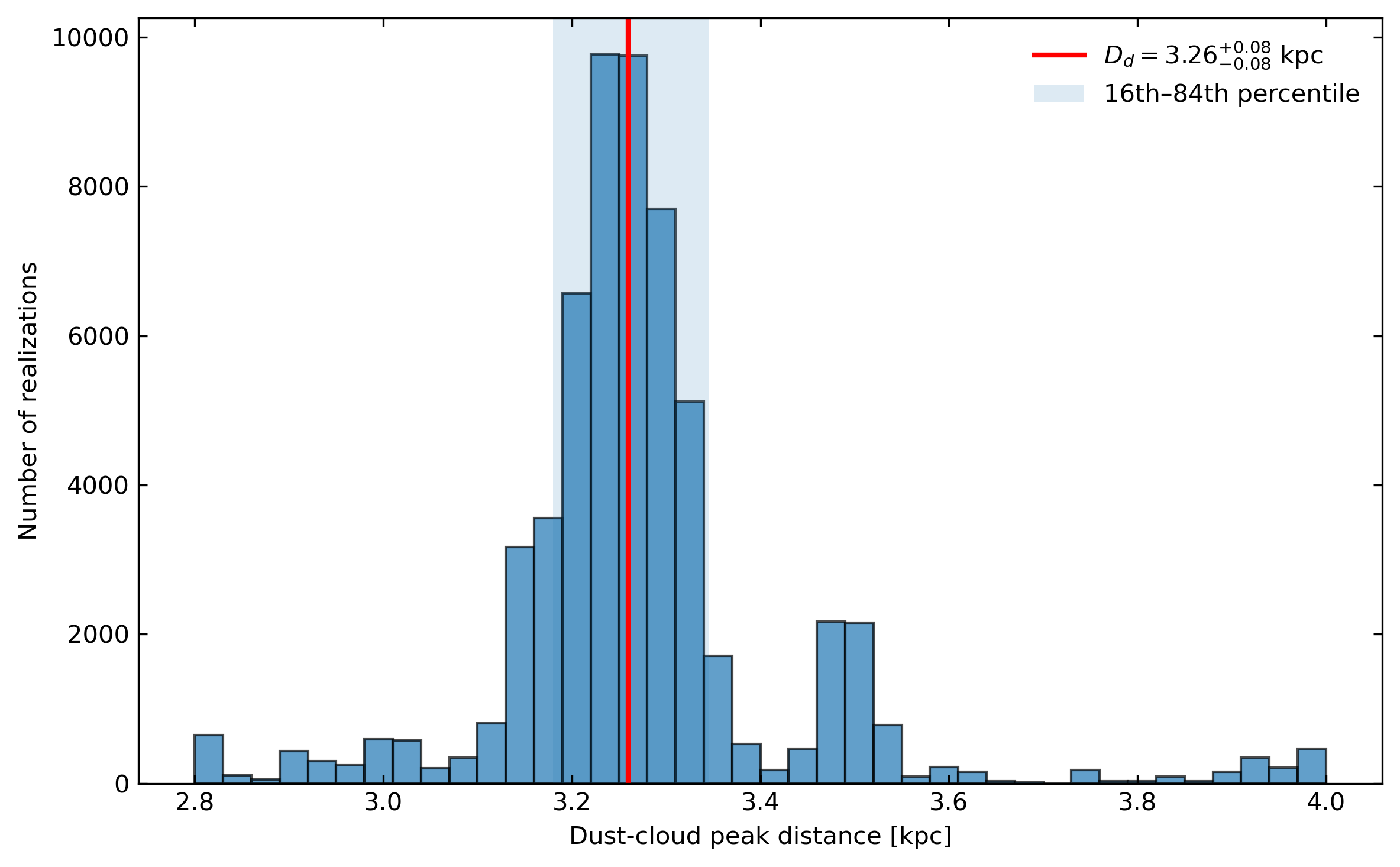}
    \caption{Bootstrap distributions of the distances of the two dust 
    clouds identified along the line of sight toward 4U~1543$-$475. 
    The distributions combine the results obtained using different 
    distance-bin widths and bin offsets. The vertical lines indicate 
    the median distances, while the shaded regions represent the 
    16th--84th percentile intervals.}
    \label{fig:boots}
\end{figure*}

To obtain a robust estimate of the dust-cloud distances and their uncertainties, we applied a bootstrap procedure to the individual stellar distance and extinction measurements from the StarHorse catalogue. The procedure was designed to account not only for the statistical fluctuations associated with the finite stellar sample, but also for the dependence of the inferred cloud position on the adopted distance binning.
For each bootstrap realization, the StarHorse stellar sample was randomly resampled with replacement and the extinction profile was reconstructed. The differential extinction profile was then derived from the variation of the median $A_V$ between consecutive distance bins and smoothed with a Gaussian kernel. The cloud distance was then identified as the position of the maximum extinction increase. We tested six different bin widths, ranging from 
0.05 to 0.20~kpc, and five different bin offsets for each bin width. 
For each binning configuration, 2000 bootstrap realizations were performed, resulting in a total of 60000 realizations for each dust cloud. 
The resulting distributions of the recovered peak positions for all 60000 realizations are shown in Fig.~\ref{fig:boots}. We adopted the median of each distribution as the cloud distance and the 16th and 84th percentiles as the lower and upper confidence limits, respectively. This yields $D_{\rm d,1}=1.84\pm0.08$~kpc and $D_{\rm d,2}=3.26\pm0.08$~kpc.

\section{Additional tracers of the interstellar medium toward 4U~1543$-$475}
\label{app:dust_los}

As a complementary check of the dust structures identified from the StarHorse catalogue, we investigated the line of sight toward 4U~1543$-$475 using the independent three-dimensional extinction map of \citet{Lallement2022} and Galactic CO emission.

\begin{figure}
\centering
\includegraphics[width=\linewidth]{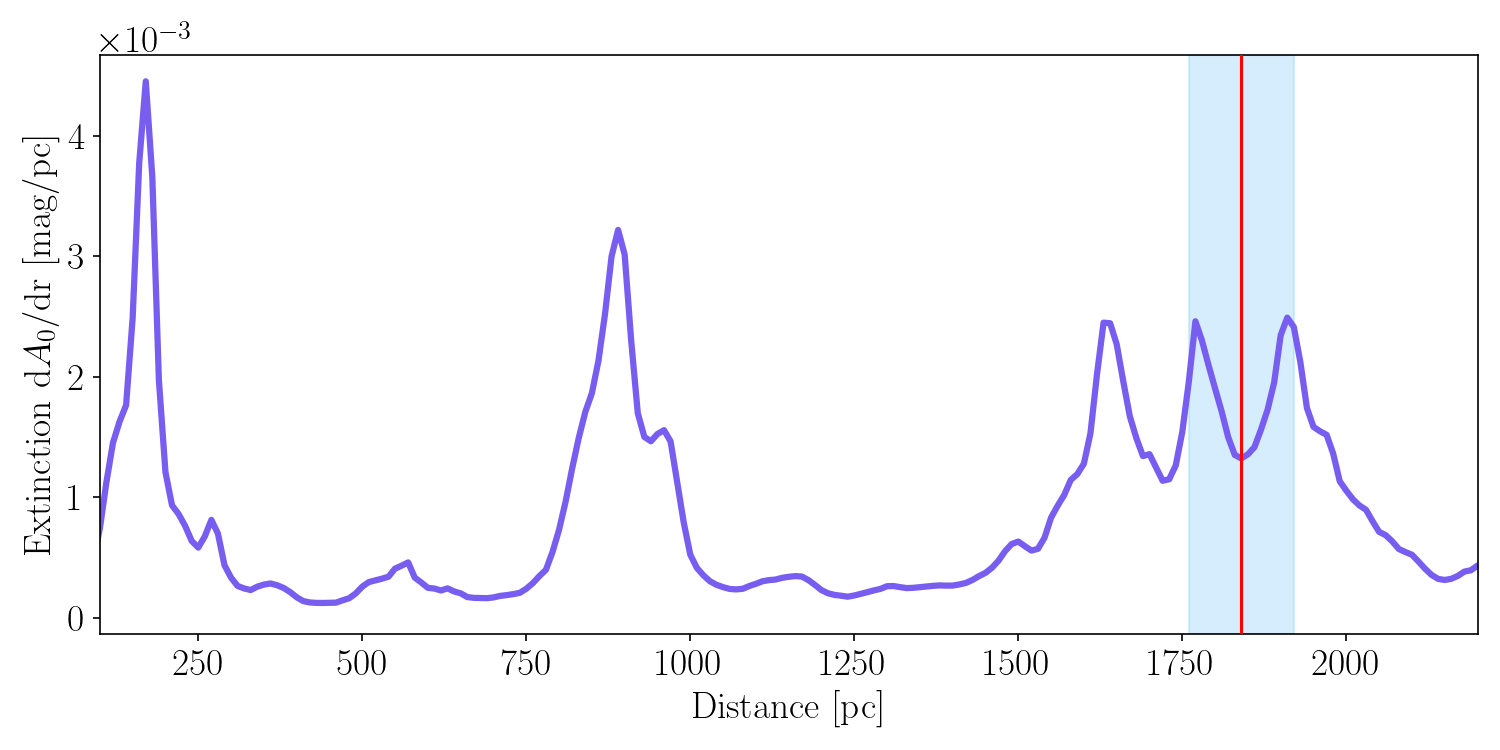}
\caption{Differential extinction along the line of sight toward 4U~1543$-$475 as a function of distance, extracted from the three-dimensional extinction map of \citet{Lallement2022}. The red vertical line indicate the dust-layer distance derived from the bootstrap analysis, while the shaded blue region represent the corresponding uncertainty.}
\label{lallement_los}
\end{figure}

The three-dimensional extinction map of \citet{Lallement2022} provides an independent reconstruction of the distribution of interstellar dust based on stellar extinction measurements. Unlike the StarHorse analysis, in which the extinction profile is reconstructed directly from individual stellar measurements, the \citet{Lallement2022} map provides a tomographic reconstruction of the differential extinction density, through a regularized Bayesian inversion. The method accounts for the irregular spatial sampling of the stellar tracers and their associated 
uncertainties.
Figure~\ref{lallement_los} shows the differential extinction along the line of sight towards 4U~1543$-$475 as a function of distance. 


\begin{figure}
\centering
\includegraphics[width=\linewidth]{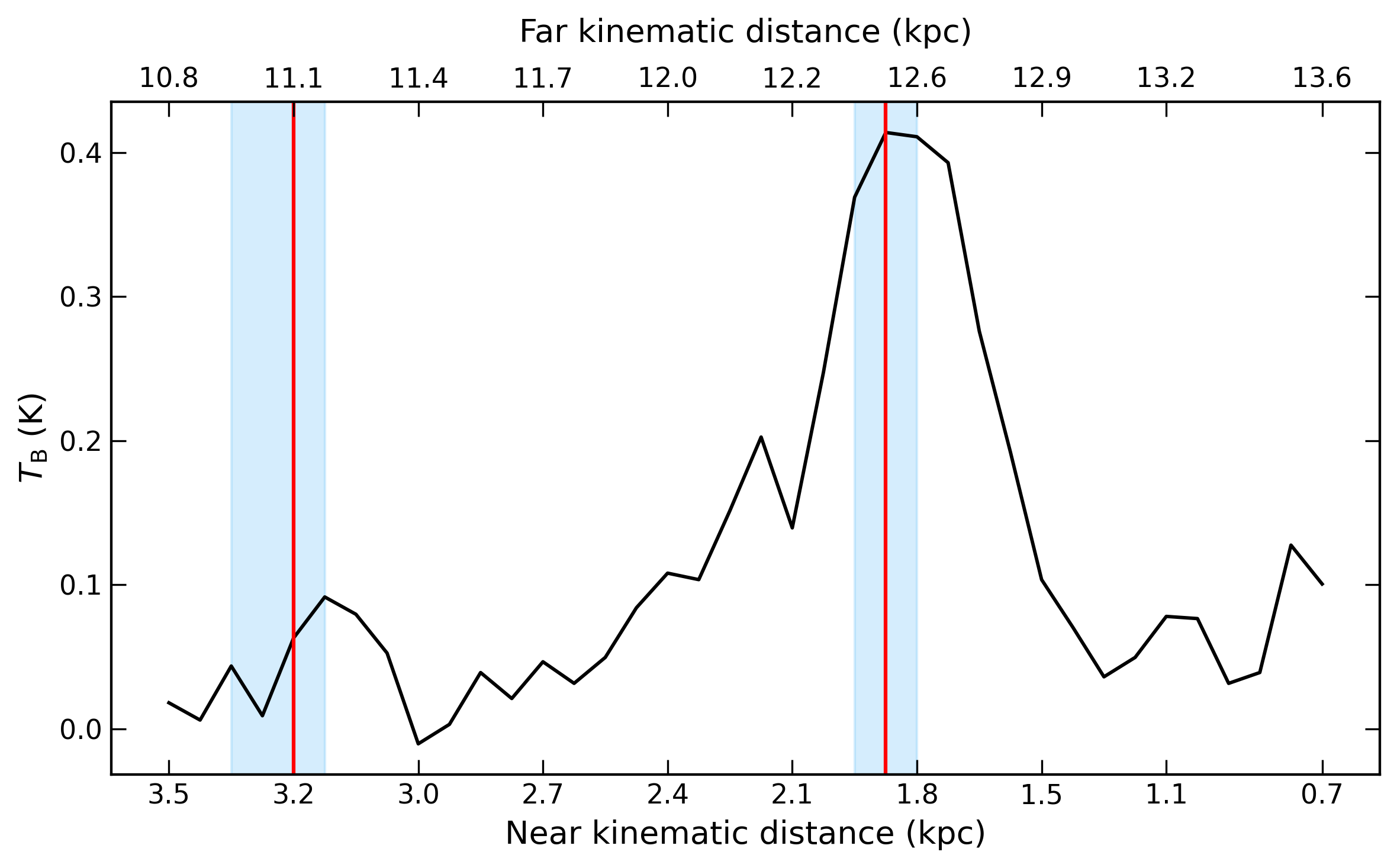}
\caption{
CO brightness-temperature spectrum, $T_{\rm B}$, along the line of 
sight toward 4U~1543$-$475, expressed as a function of the kinematic 
distance derived using the Galactic rotation curve of 
\citet{Reid2019}. The lower and upper axes show the near and far 
kinematic-distance solutions, respectively. The red vertical lines indicate the dust-layer distances derived from the bootstrap analysis, while the shaded blue regions represent their corresponding uncertainties.
}
\label{co_los}
\end{figure}

The Galactic CO emission \citep{Dame2001,Dame2011} provides an independent tracer of the molecular gas distribution along the line of sight. The CO emission is measured as a function of the local 
standard of rest (LSR) velocity, which we converted into kinematic distances using the Galactic rotation curve of \citet{Reid2019}. This model, based on trigonometric parallaxes and proper motions of masers associated with high-mass star-forming regions, adopts a Galactocentric distance of the Sun of $R_0=8.15$~kpc and a circular rotation velocity of $\Theta_0=236$~km~s$^{-1}$.
Because a given radial velocity generally corresponds to two possible heliocentric distances, we computed both the near and far kinematic distance solutions. The resulting CO distribution is shown in Fig.~\ref{co_los}. 

The molecular-gas distribution shows structures at distances consistent with the two extinction enhancements identified using StarHorse and used to constrain the distance of 4U~1543$-$475.
Given the dependence of the CO-derived distances on the adopted Galactic rotation curve, as well as the near--far kinematic distance ambiguity, we did not use the CO emission to quantitatively determine the cloud distances. Instead, the CO data are used here as an independent tracer supporting the presence of interstellar structures along the line of sight.

The molecular-gas distribution shows structures at distances consistent with both extinction enhancements identified using StarHorse, at approximately 1.8 and 3.3~kpc, whereas the \citet{Lallement2022} extinction map only traces the nearer dust layer, the other one is beyond the map limit.

\bibliography{main}
\bibliographystyle{aa}

\end{document}